\documentclass[conference]{IEEEtran}
\usepackage{cite}

\ifCLASSINFOpdf
\usepackage[pdftex]{graphicx}
\else
\fi
\usepackage{amsmath}
\usepackage{amsfonts} 
\usepackage{amsthm}
\usepackage{amssymb}
\usepackage[boxruled]{algorithm2e}

\usepackage{BOONDOX-ds}

\usepackage{bbm}

\usepackage{kbordermatrix}

\newcommand{\pfun}{\mathop{\hbox{$\to$\kern-7pt\raise.9pt\hbox{\scalebox{1}[.55]{$|$}}\kern4pt} }}

\usepackage{array}

\begin{document}

\title{GraphChallenge.org: Raising the Bar on Graph Analytic Performance}

\author{\IEEEauthorblockN{Siddharth Samsi,
Vijay Gadepally,
Michael Hurley,
Michael Jones,
Edward Kao,
Sanjeev Mohindra, \\
Paul Monticciolo,
Albert Reuther,
Steven Smith,
William Song,
Diane Staheli,
Jeremy Kepner \\
\IEEEauthorblockA{MIT Lincoln Laboratory, Lexington, MA}}}
\maketitle

\begin{abstract}
The rise of graph analytic systems has created a need for new ways to measure and compare
the capabilities of graph processing systems. The MIT/Amazon/IEEE Graph Challenge has been developed to provide a well-defined community venue for stimulating research and highlighting innovations in graph analysis software, hardware, algorithms, and systems.  GraphChallenge.org provides a wide range of pre-parsed graph data sets, graph generators, mathematically defined graph algorithms, example serial implementations in a variety of languages, and specific metrics for measuring performance.  Graph Challenge 2017 received 22 submissions by 111 authors from 36 organizations.  The submissions highlighted graph analytic innovations in hardware, software, algorithms, systems, and visualization.  These submissions produced many comparable performance measurements that can be used for assessing the current state of the art of the field.  There were numerous submissions that implemented the triangle counting challenge and resulted in over 350 distinct measurements.  Analysis of these submissions show that their execution time is a strong function of the number of edges in the graph, $N_e$, and is typically proportional to $N_e^{4/3}$ for large values of $N_e$.  Combining the model fits  of the submissions presents a picture of the current state of the art of graph analysis, which is typically $10^8$ edges processed per second for graphs with $10^8$ edges. These results are $30$ times faster than serial implementations commonly used by many graph analysts and underscore the importance of making these performance benefits available to the broader community.  Graph Challenge provides a clear picture of current graph analysis systems and underscores the need for new innovations to achieve high performance on very large graphs.
\end{abstract}

%
\IEEEpeerreviewmaketitle

\section{Introduction}
\let\thefootnote\relax\footnotetext{This material is based upon work supported by the Assistant Secretary of Defense for Research and Engineering under Air Force Contract No. FA8721-05-C-0002 and/or FA8702-15-D-0001. Any opinions, findings, conclusions or recommendations expressed in this material are those of the author(s) and do not necessarily reflect the views of the Assistant Secretary of Defense for Research and Engineering.}

The importance of graph analysis has dramatically increased and is critical to a wide variety of domains that include the analysis of genomics \cite{Morrison2005,Mooney2012,polychronopoulos2014conserved,Dodson2014,Dodson2015,gouda2016distribution}, brain mapping \cite{fornito2016graph}, computer networks \cite{BrinPage1998,Faloutsos1999,yan2015spectrum,fontugne2017scaling}, social media \cite{Zuckerburg2005,Kwak2009}, cybersecurity \cite{shao2015percolation,yu2015malware}, and sparse machine learning \cite{lee2008sparse,boureau2008sparse,glorot2011deep,yu2012exploiting,Kepner2017graphblasDNN}.

  Many graph processing systems are currently under development.  These systems are exploring innovations in algorithms \cite{Cormen2001,Miller2012,Buluc2014,voegele2017parallel,smith2017truss,hu2017trix,la2017ensemble,zhuzhunashvili2017preconditioned,low2017first,uppal2017scalable,mowlaei2017triangle}, software architecture \cite{BulucGilbert2011,Kepner2012-ch1,pearce2017triangle,halappanavar2017scalable,tom2017exploring,green2017quickly,kabir2017parallel,zhou2017design,hutchison2017distributed,wolf2017fast}, software standards \cite{Mattson2013,kepner2015graphs,kepner2016mathematical,bulucc2017design,davis2017graphblas}, and parallel computing hardware \cite{song2016novel,bisson2017static,date2017collaborative,debenedictis2017superstrider,manne2017if,kogge2017graph,gioiosa2017exploring}.
The rise of graph analysis systems has created a need for new ways to measure and compare
the capabilities of these systems. The MIT/Amazon/IEEE Graph Challenge has been developed to provide a well-defined community venue for stimulating research and highlighting innovations in graph analysis software, hardware, algorithms, and systems.   
GraphChallenge.org provides a wide range of pre-parsed graph data sets, graph generators, mathematically defined graph algorithms, example serial implementations in a variety of languages, and specific metrics for measuring performance.

Scale is an important driver of the Graph Challenge and graphs with billions to
trillions of edges are of keen interest.  The Graph Challenge is designed to work on
arbitrary graphs drawn from both real-world data sets and simulated data sets.
Examples of real-world data sets include the Stanford Large Network Dataset
Collection (see http://snap.stanford.edu/data), the AWS Public Data Sets (see
aws.amazon.com/public-data-sets), and the Yahoo! Webscope Datasets (see
webscope.sandbox.yahoo.com).  These real-world data sets cover a wide range of
applications and data sizes.  While real-world data sets have many contextual
benefits, synthetic data sets allow the largest possible graphs to be readily
generated. Examples of synthetic data sets include Graph500, Block Two-level
Erdos-Renyi graph model (BTER) \cite{seshadhri2012community}, Pure Kronecker Graphs
\cite{kepner2011graph}, and  Perfect Power Law graphs
\cite{kepner2012perfect,gadepally2015using}. The focus of the Graph Challenge is on graph analytics.  While parsing and formatting
complex graph data are necessary in any graph analysis system, these data sets are
made available to the community in a variety of pre-parsed formats to minimize the
amount of parsing and formatting  required by Graph Challenge participants.  The public data
are available in a variety of formats, such as linked list, tab separated, and
labeled/unlabeled.

Graph Challenge 2017 received a large number of submissions that highlighted innovations in hardware, software, algorithms, systems, and visualization.  These submissions produced many comparable performance measurements that can be used for assessing the current state of the art of the field.  The goal of this paper is to analyze and synthesize these measurements to provide a picture of the current state of the art of graph analysis systems.  The organization of this paper is as follow.  First, a recap of triangle counting is provided, along with a few standard algorithms.  Next, an overview is presented of the Graph Challenge 2017 submissions.  The core of the paper is the section on the analysis of the 12 submission that all performed the triangle counting challenge.  Based on this analysis, these results are synthesized to provide a picture of the current state of the art.  Finally, the conclusions are provided along with a discussion of further work.

\section{Triangle Counting}

The Graph Challenge consists of three challenges
\begin{itemize}
\item Pre-challenge: PageRank pipeline \cite{dreher2016pagerank}
\item Static graph challenge: subgraph isomorphism \cite{samsi2017static}
\item Streaming graph challenge: stochastic block partition \cite{ed}
\end{itemize}
The static graph challenge is further broken down into triangle counting and k-truss.  Triangle counting received the most submissions and is the focus of this paper.

\begin{figure}[ht]
\centering
\includegraphics[width=\columnwidth]{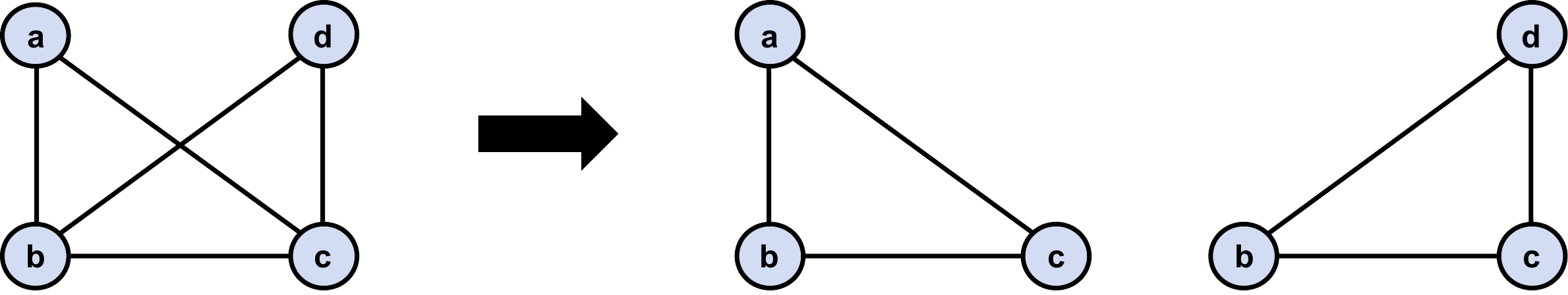}
\caption{The graph shown in this example contains two triangles consisting of nodes \{a,b,c\} and \{b,c,d\}.}
\label{fig:triangle}
\end{figure}

Triangles are the most basic, trivial sub-graph. A triangle can be defined as a
set of three mutually adjacent vertices in a graph. As shown in Figure~\ref{fig:triangle}, the graph \textbf{G} contains two triangles comprising nodes \{a,b,c\} and \{b,c,d\}. The number of triangles in a graph
is an important metric used in applications such as social network mining, link
classification and recommendation, cyber security, functional biology, and spam
detection ~\cite{pavan2013}.

The number of triangles in a given graph \textbf{G} can be calculated in several
ways. We highlight two algorithms based on linear algebra primitives. The first
algorithm proposed by Wolf et al~\cite{wolf2015} uses an overloaded matrix
multiplication approach on the adjacency and incidence matrices of the graph and is
shown in Algorithm~\ref{triangle_1}. The second approach proposed by Burkhardt
et al~\cite{burkhardt2016} uses only the adjacency matrix of the given graph and is
shown in Algorithm~\ref{triangle_2}.

\begin{algorithm}[h]
  \KwData{Adjacency matrix \textbf{A} and incidence matrix \textbf{E}}
  \KwResult{Number of triangles in graph \textbf{G}}
  initialization\;
  $\textbf{C} = \textbf{A}\textbf{E}$\\
  $n_{T} = nnz(C)/3$\\
  \vspace{.25cm}
  Multiplication is overloaded such that \\
  $\textbf{C}(i,j) = \{i, x, y\}$ iff \\
  $\textbf{A}(i,x) = \textbf{A}(i,y) = 1$ \& $\textbf{E}(x,j) = \textbf{E}(y,j) = 1$ \\
  \vspace{.2cm}
  \caption{Array based implementation of triangle counting algorithm using the adjacency and incidence matrix of a graph~\cite{wolf2015}.}
  \label{triangle_1}
\end{algorithm}

\begin{algorithm}[ht]
\KwData{Adjacency matrix \textbf{A}}
\KwResult{Number of triangles in graph \textbf{G}}
initialization\;
$\textbf{C} = \textbf{A}^2 \circ \textbf{A}$ \\
$n_{T} = \sum_{ij}^{ } (\textbf{C}) / 6$ \\
\vspace{1em}
Here, $\circ$ denotes element-wise multiplication\\
\vspace{.2cm}
\caption{Array based implementation of triangle counting algorithm using only the adjacency matrix of a graph~\cite{burkhardt2016}.}
\label{triangle_2}
\end{algorithm}

Another algorithm for triangle counting based on a masked matrix multiplication
approach has been proposed by Azad et al~\cite{gilbert2015}. The serial version of
this algorithm based on the MapReduce implementation by Cohen et al~\cite{cohen2009}
is shown in Algorithm~\ref{triangle_3}. Finally, a comparison of triangle counting
algorithms can be found in~\cite{wang2016}.

\begin{algorithm}[h]
\KwData{Adjacency matrix \textbf{A}}
\KwResult{Number of triangles in graph \textbf{G}}
initialization\;
$(\textbf{L}, \textbf{U}) \leftarrow \textbf{A}$\\
$\textbf{B} = \textbf{L}\textbf{U} $ \\
$\textbf{C} = \textbf{A} \circ \textbf{B} $ \\
$n_{T} = \sum_{ij}^{ } (\textbf{C}) / 2$ \\
\vspace{1em}
Here, $\circ$ denotes element-wise multiplication\\
\vspace{.2cm}
\caption{Serial version of triangle counting algorithm based on MapReduce version by Cohen et al~\cite{cohen2009} and ~\cite{gilbert2015}.}
\label{triangle_3}
\end{algorithm}

\section{Community Submissions}

\begin{table}
\caption{{\rm Breakdown of the Graph Challenge 2017 submissions amongst organizations, challenges, innovations, and awards.}}
\centering
\begin{tabular}{cccc}
& {\bf Organizations} \\
\hline
Academic & Federal & Industry \\
\hline
     20  &     9   &    7    \\
\hline
\\
\end{tabular}
\begin{tabular}{ccc}
& {\bf Challenge} \\
\hline
Static & Streaming & PageRank \\
\hline
     17  &   3   &    1       \\
\hline
\\
\end{tabular}
\begin{tabular}{ccc}
& {\bf Static Kernel} \\
\hline
Triangle & K-Truss & Both \\
\hline
      10  &  3   &   4   \\
\hline
\\
\end{tabular}
\begin{tabular}{ccccc}
& & {\bf Innovation} \\
\hline
Hardware & Software & Algorithms & Viz & Systems \\
\hline
      5  &  13      &     13     &  1  &  14 \\
\hline
\\
\end{tabular}
\begin{tabular}{ccccc}
& & {\bf Award} \\
\hline
Champion & Finalist & Innovation & Student    & Honorable \\
         &          &            & Innovation & Mention \\
\hline
      5  &  13      &     13     &  1  &  14 \\
\hline
\end{tabular}
\label{table:breakdown}
\end{table}

Graph Challenge 2017 received 22 submissions by 111 authors from 36 organizations \cite{voegele2017parallel,smith2017truss,hu2017trix,la2017ensemble,zhuzhunashvili2017preconditioned,low2017first,uppal2017scalable,mowlaei2017triangle,pearce2017triangle,halappanavar2017scalable,tom2017exploring,green2017quickly,kabir2017parallel,zhou2017design,hutchison2017distributed,wolf2017fast,bisson2017static,date2017collaborative,debenedictis2017superstrider,manne2017if,kogge2017graph,gioiosa2017exploring}. The submissions were judged by a panel of experts on their effectiveness at using Graph Challenge to highlight innovations in graph algorithms, hardware, software, and systems. Because of the technical expertise required to submit, submitting successfully is a significant accomplishment, and potential participants who were unable to effectively complete the challenge simply did not submit.  The breakdown of the submissions amongst organizations, challenges, innovations, and awards is shown in Table~\ref{table:breakdown}.  

\begin{figure}[ht]
\centering
\includegraphics[width=\columnwidth]{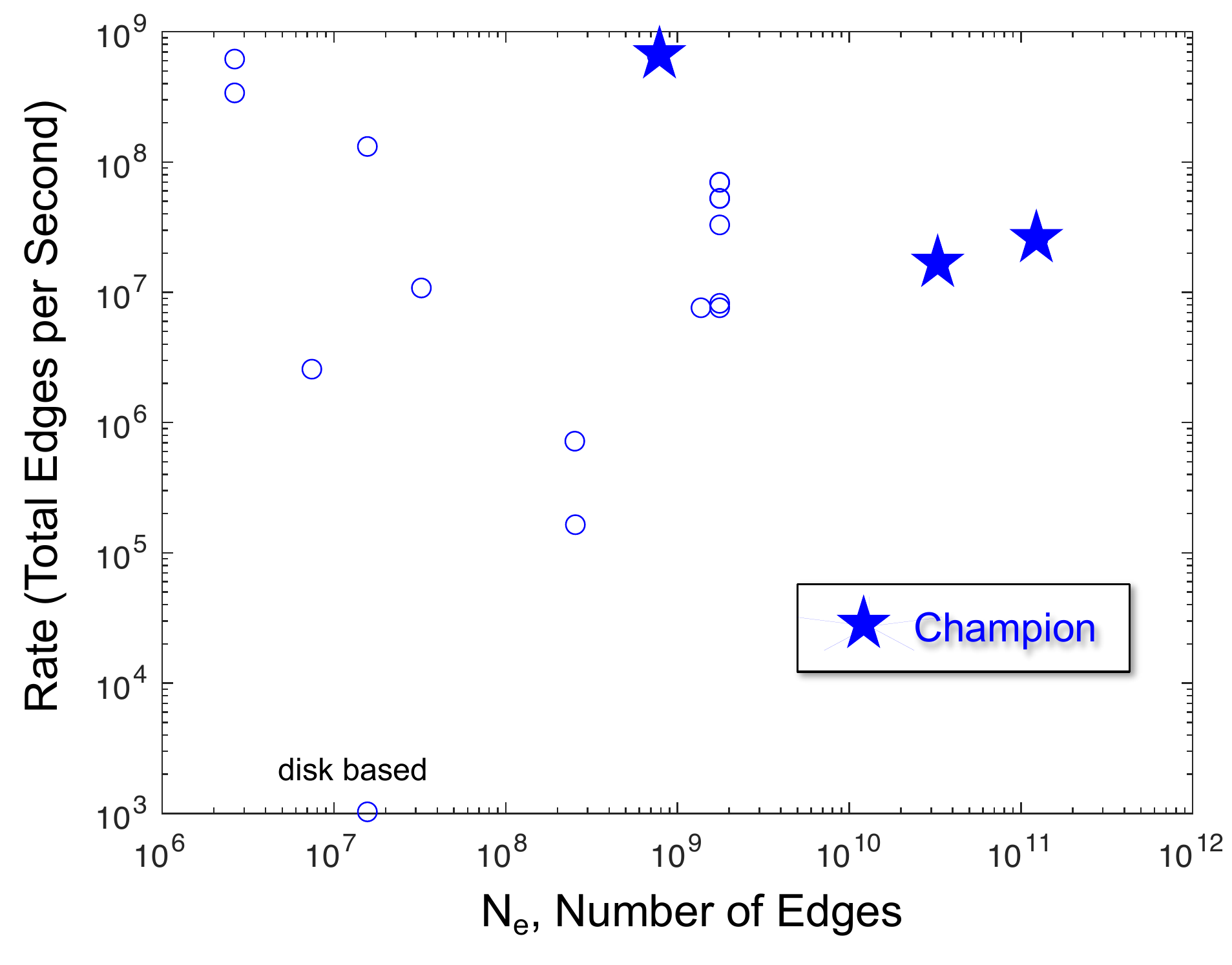}
\caption{The largest graph and corresponding rate and the fastest graph and corresponding size for selected Graph Challenge 2017 triangle counting submissions.  The one disk-based submission shows the importance of large, high-speed memories for this application.}
\label{fig:PerformanceSummary}
\end{figure}

Numerous submissions implemented the triangle counting challenge in a comparable manner,  resulting in over 350 distinct measurements of triangle counting execution time, $T_{\rm tri}$.  The number of edges, $N_e$, in the graph describes the overall size of the graph.  The rate of edges processed in triangle counting is given by
$$
 {\rm Rate} = N_e/T_{\rm tri}
$$
For any of these submissions, it is possible to extract two points: (1) the highest rate and its corresponding number of edges, and (2) the largest graph and its corresponding rate.  These points are shown in Figure~\ref{fig:PerformanceSummary} and give a rough picture of the overall landscape of these submissions and highlight that the judges preferred submissions that showed high performance on large graphs.  By analyzing all the data points, it is possible to obtain a detailed picture of the state of the art of this field.

\section{Performance Analysis}

\begin{figure}[ht]
\centering
\includegraphics[width=2.5in]{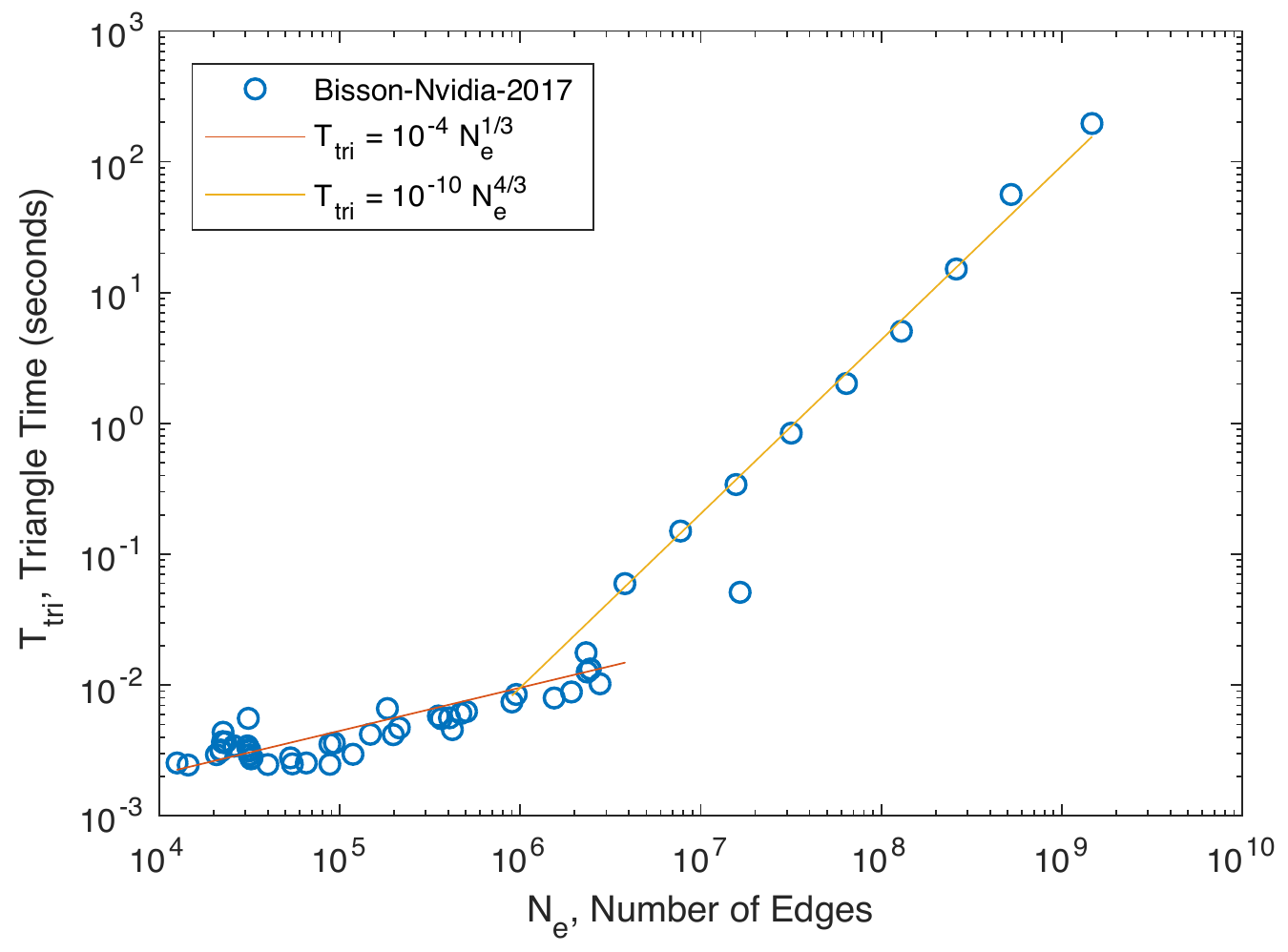}
\includegraphics[width=2.5in]{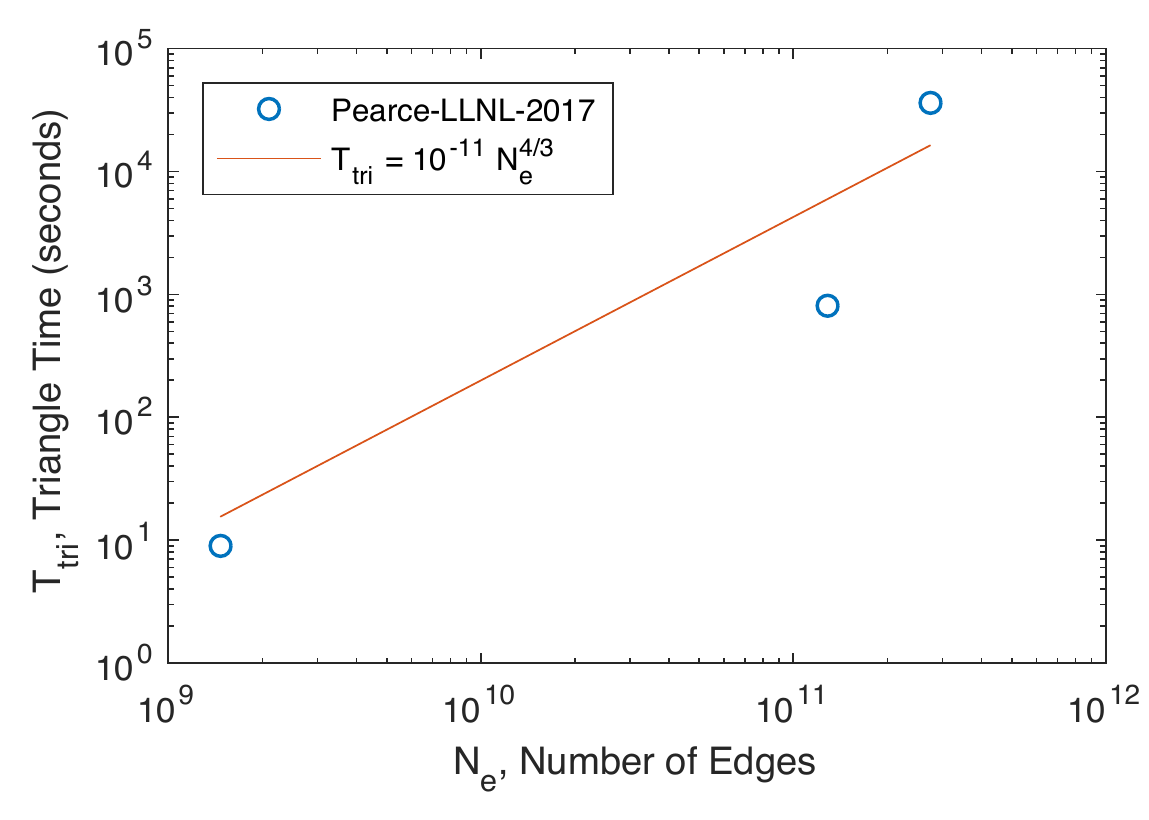}
\includegraphics[width=2.5in]{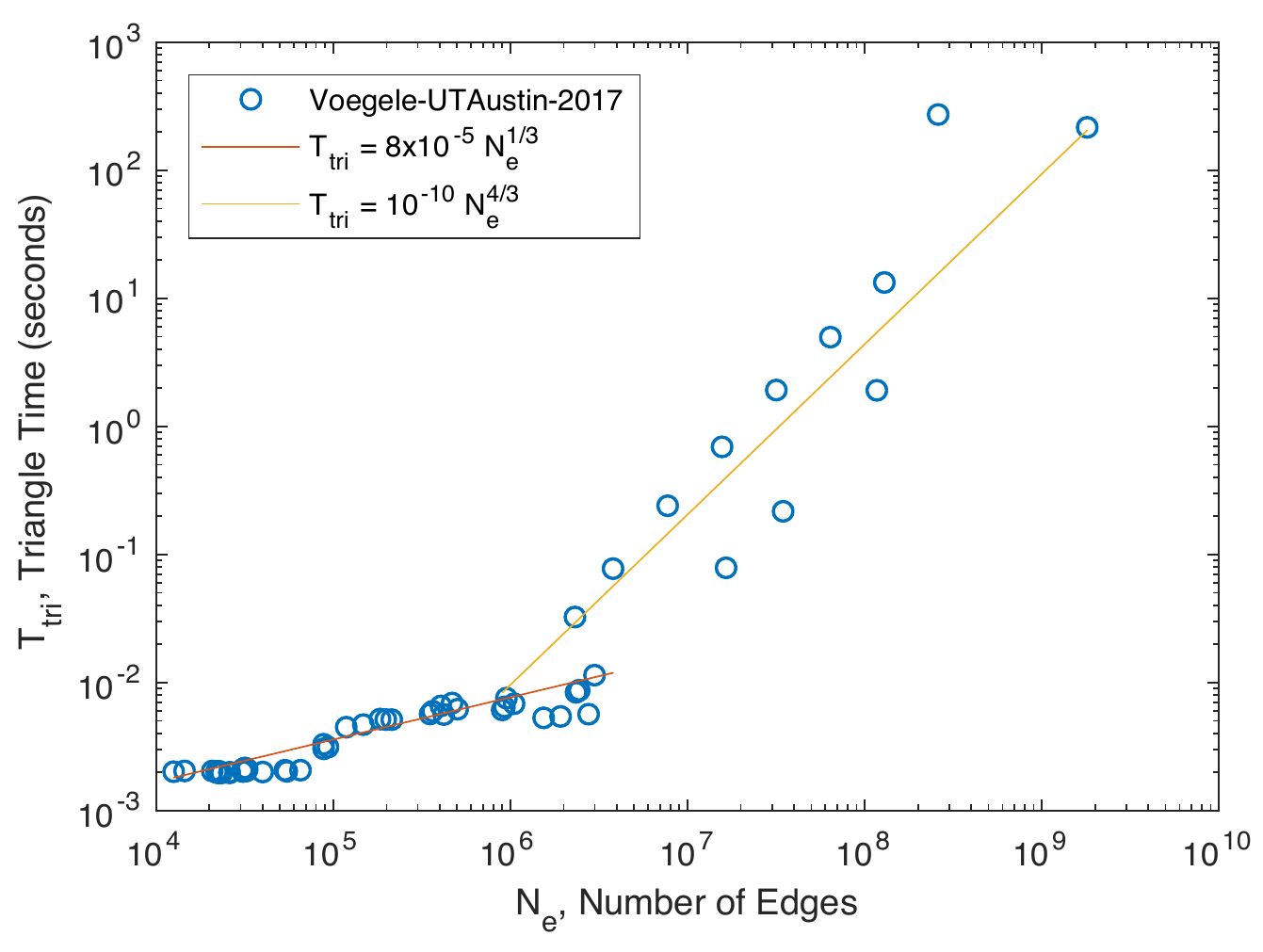}
\includegraphics[width=2.5in]{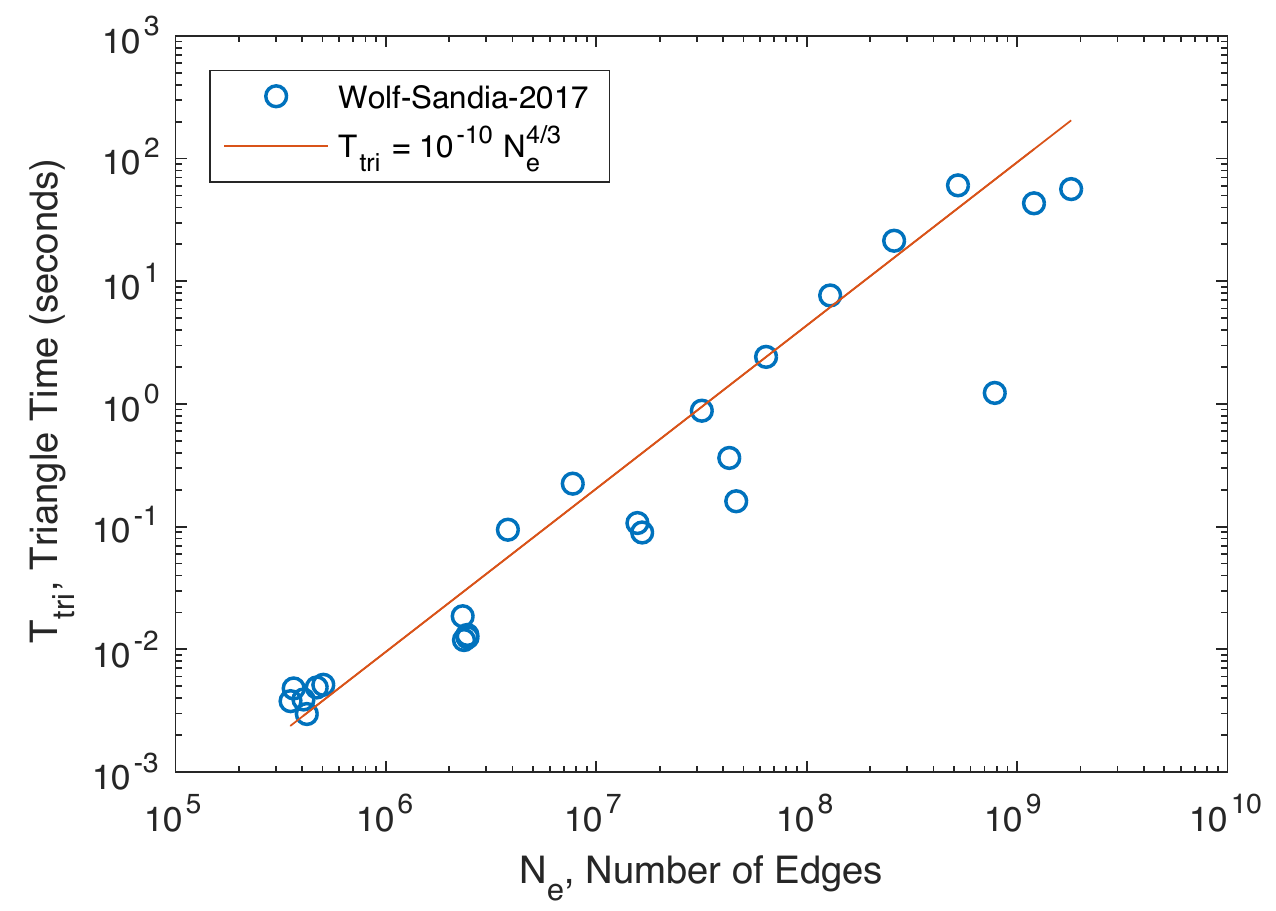}
\caption{Graph Challenge 2017 Champions. Triangle counting execution time vs number of edges and corresponding model fits for Bisson-Nvidia-2017 \cite{bisson2017static}, Pearce-LLNL-2017 \cite{pearce2017triangle}, Voegele-UTAustin-2017 \cite{voegele2017parallel}, and  Wolf-Sandia-2017 \cite{wolf2017fast}.}
\label{fig:Champions}
\end{figure}

\begin{figure}[ht]
\centering
\includegraphics[width=2.5in]{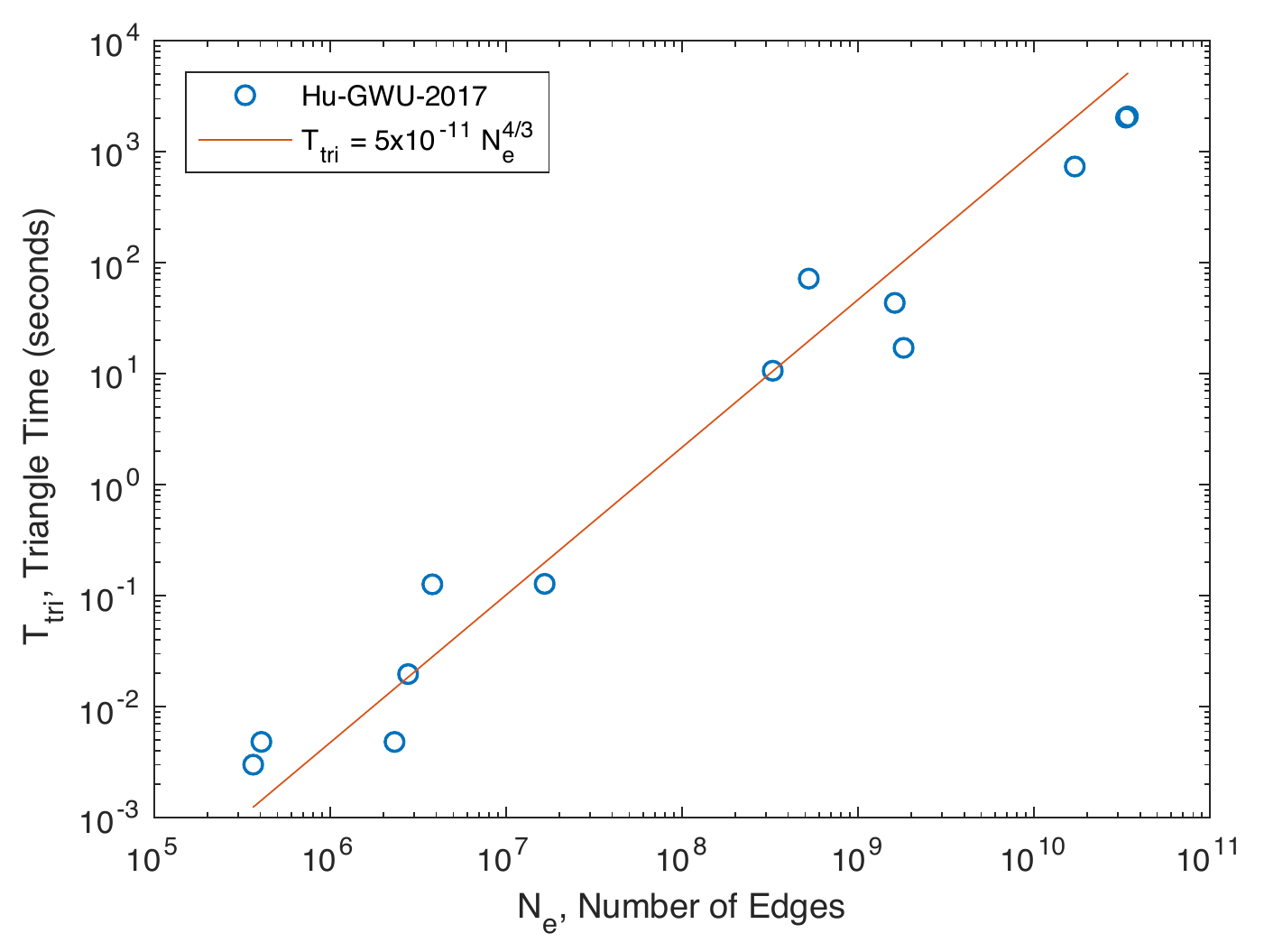}
\includegraphics[width=2.5in]{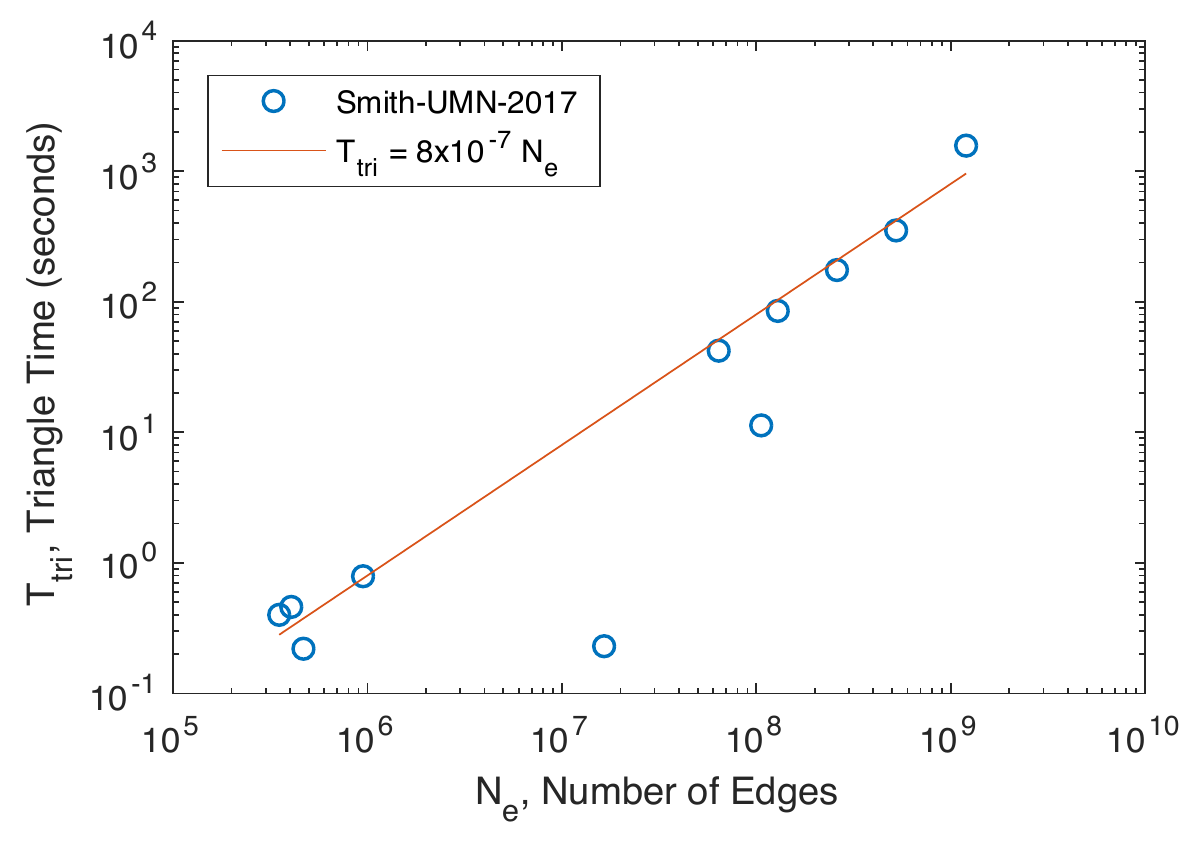}
\includegraphics[width=2.5in]{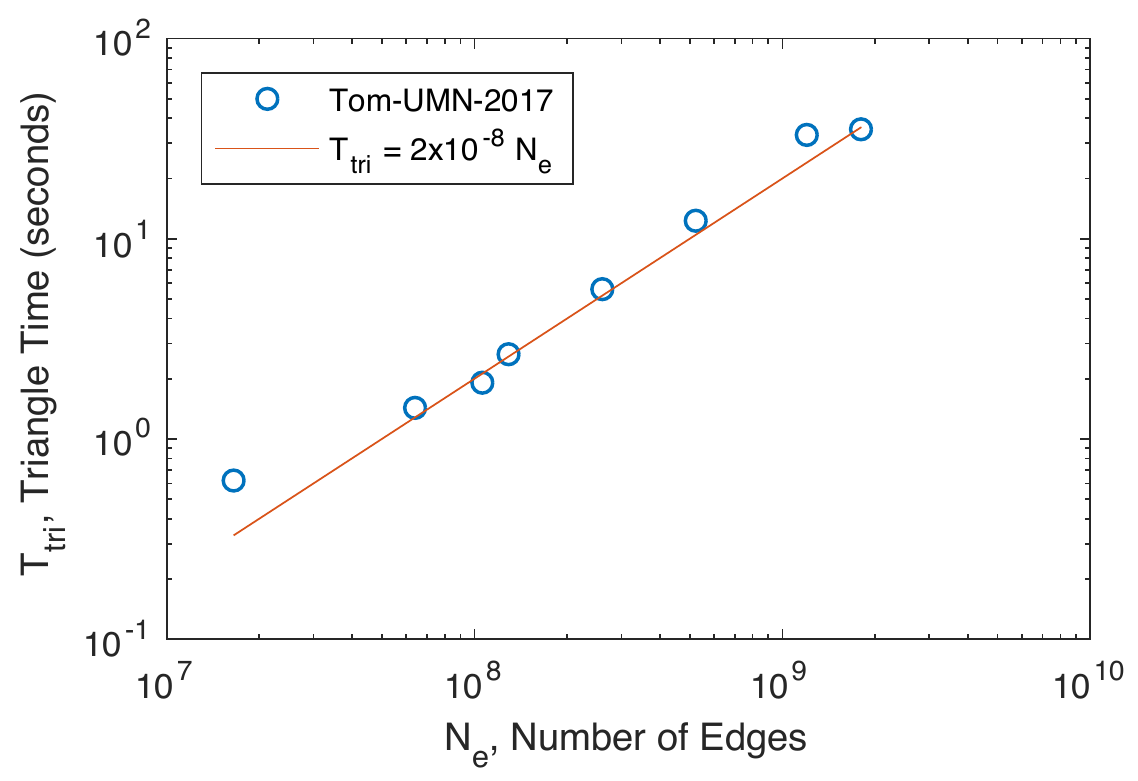}
\caption{Graph Challenge 2017 Finalists. Triangle counting execution time vs number of edges and corresponding model fits for Hu-GWU-2017-2017 \cite{hu2017trix}, Smith-UMN-2017 \cite{smith2017truss}, and Tom-UMN-2017-2017 \cite{tom2017exploring}.}
\label{fig:Finalists}
\end{figure}

\begin{figure}[ht]
\centering
\includegraphics[width=2.5in]{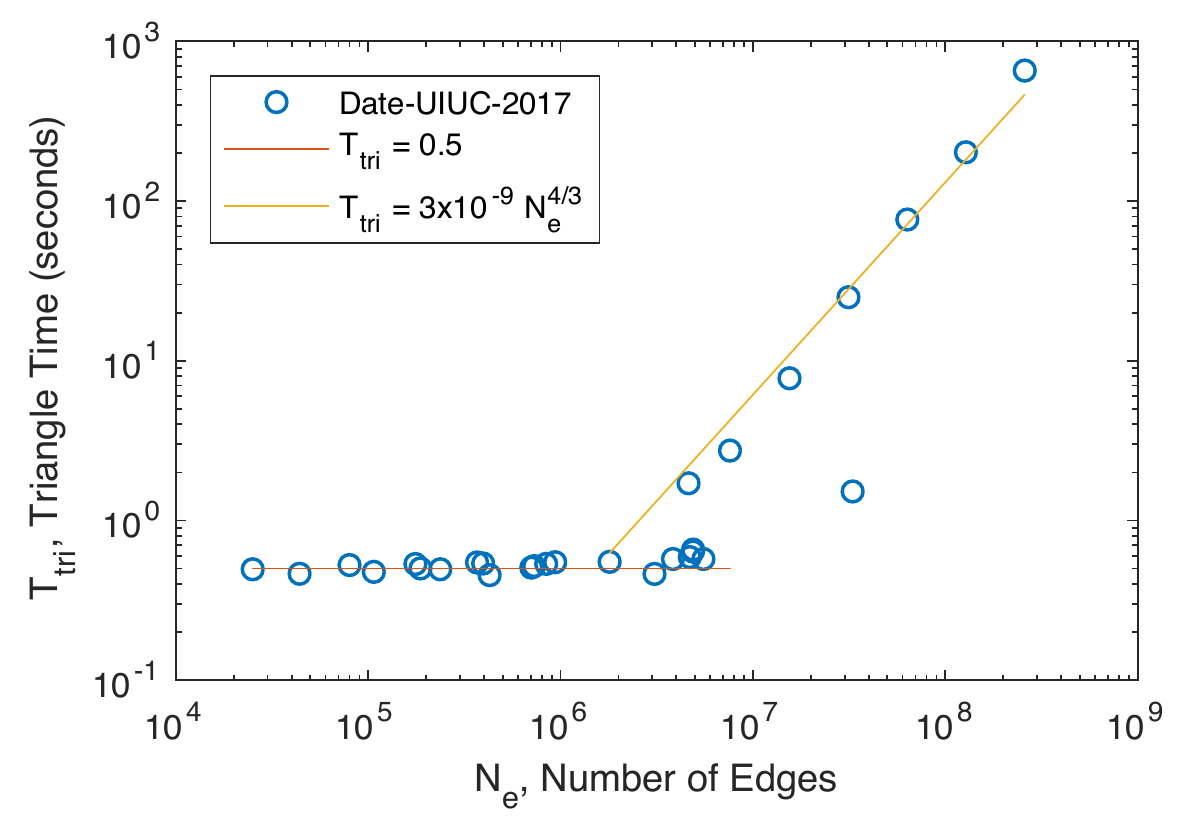}
\includegraphics[width=2.5in]{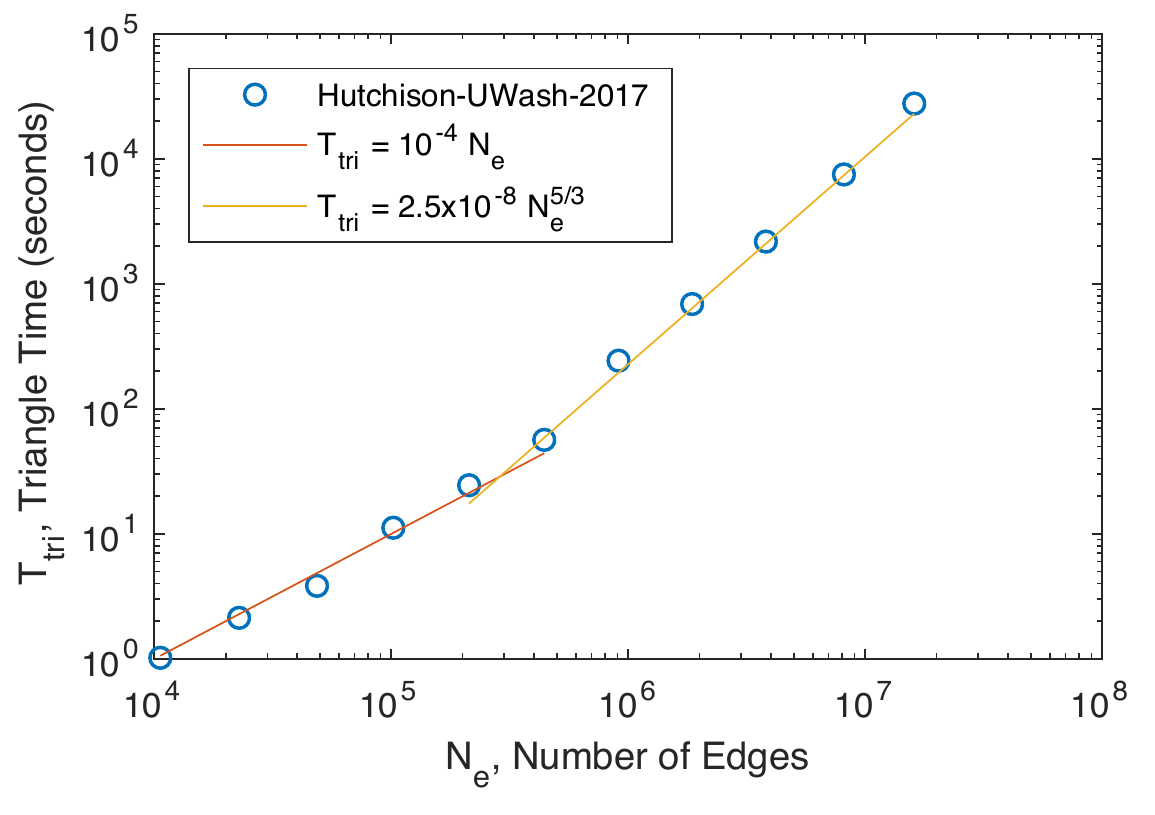}
\includegraphics[width=2.5in]{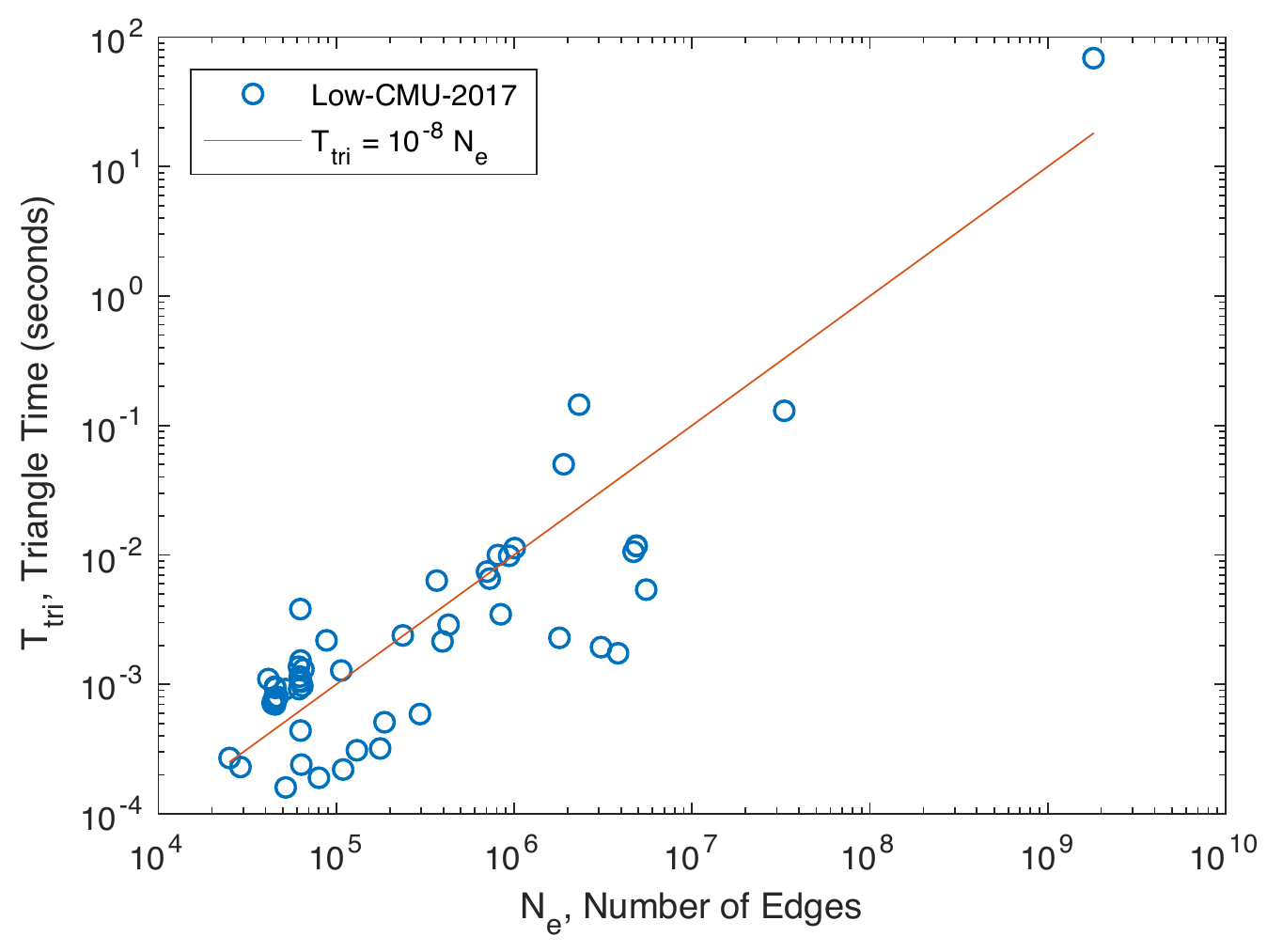}
\includegraphics[width=2.5in]{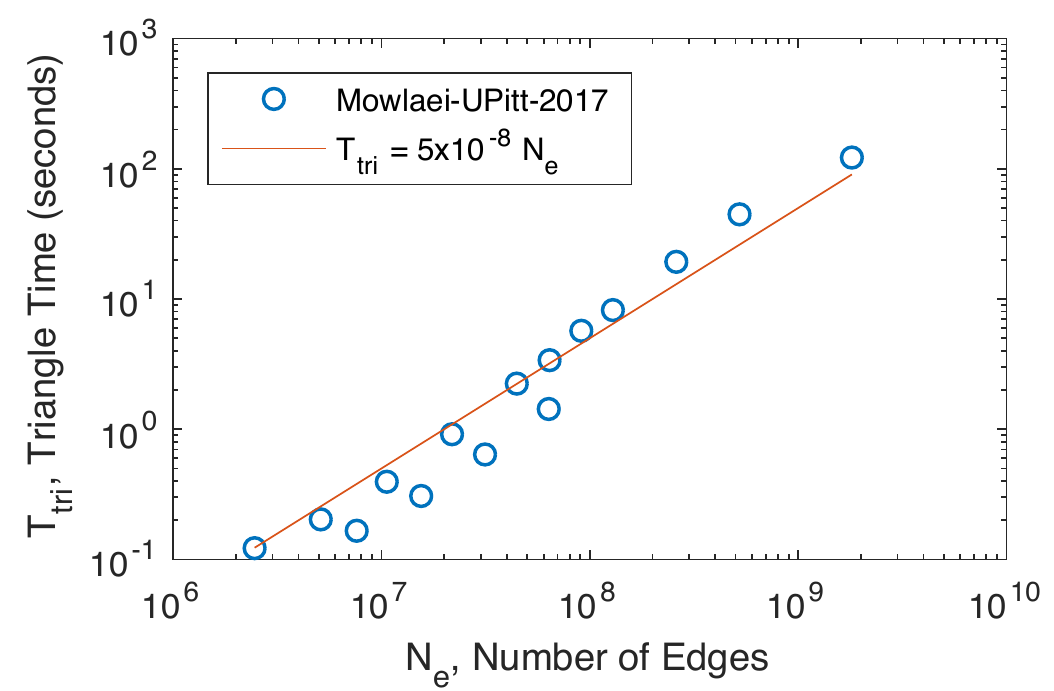}
\caption{Graph Challenge 2017 Honorable Mentions. Triangle counting execution time vs number of edges and corresponding model fits for
Date-UIUC-2017 \cite{date2017collaborative}, Hutchison-UWash-2017 \cite{hutchison2017distributed}, Low-CMU-2017 \cite{low2017first}, and Mowlaei-UPitt-2017 \cite{mowlaei2017triangle}.}
\label{fig:HonorableMentions}
\end{figure}

Analyzing and combining all the performance data from the submissions can be done by fitting a piecewise model to each submission and then comparing the models.  For each submission, $T_{\rm tri}$ vs $N_e$ is plotted on a log-log scale from which a  model can be fit to the data by estimating the parameters $\alpha$ and $\beta$ in the formula
$$
   T_{\rm tri} = \alpha N_e^\beta
$$
The triangle counting execution time vs number of edges and corresponding model fits are shown for the Champions in Figure~\ref{fig:Champions}, the Finalists in Figure~\ref{fig:Finalists}, and the Honorable Mentions in Figure~\ref{fig:HonorableMentions}.  The model fits illustrate the strong dependence of $T_{\rm tri}$ on $N_e$.   

\section{Performance Synthesis}

\begin{figure}[ht]
\centering
\includegraphics[width=\columnwidth]{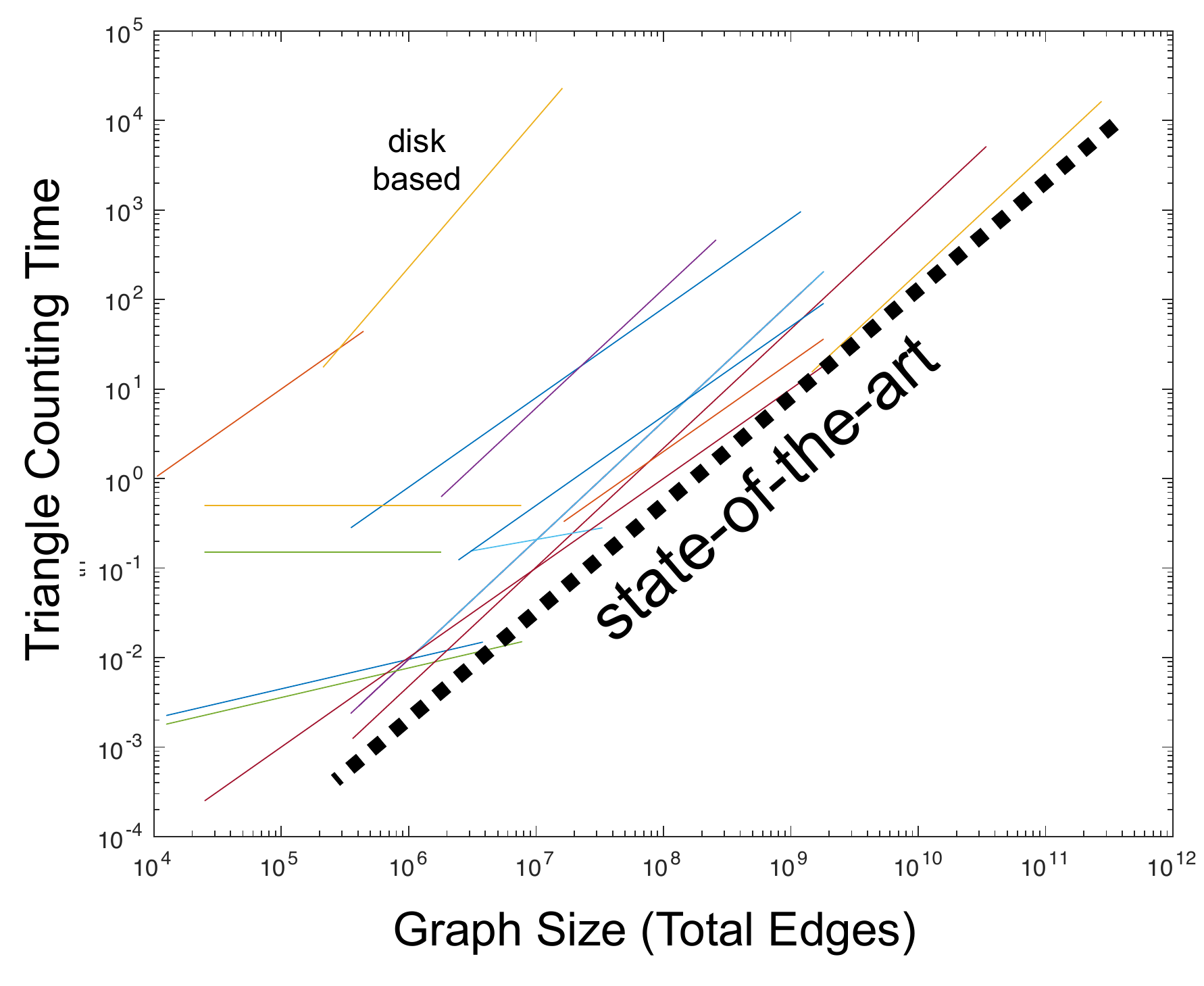}
\caption{Model fits of triangle execution time vs number edges for selected Graph Challenge 2017 triangle counting submissions.  State-of-the-art is denoted by the black dashed line. The one disk-based submissions shows the importance of large, high-speed memories for this application.}
\label{fig:TimePerformance}
\end{figure}

\begin{figure}[ht]
\centering
\includegraphics[width=\columnwidth]{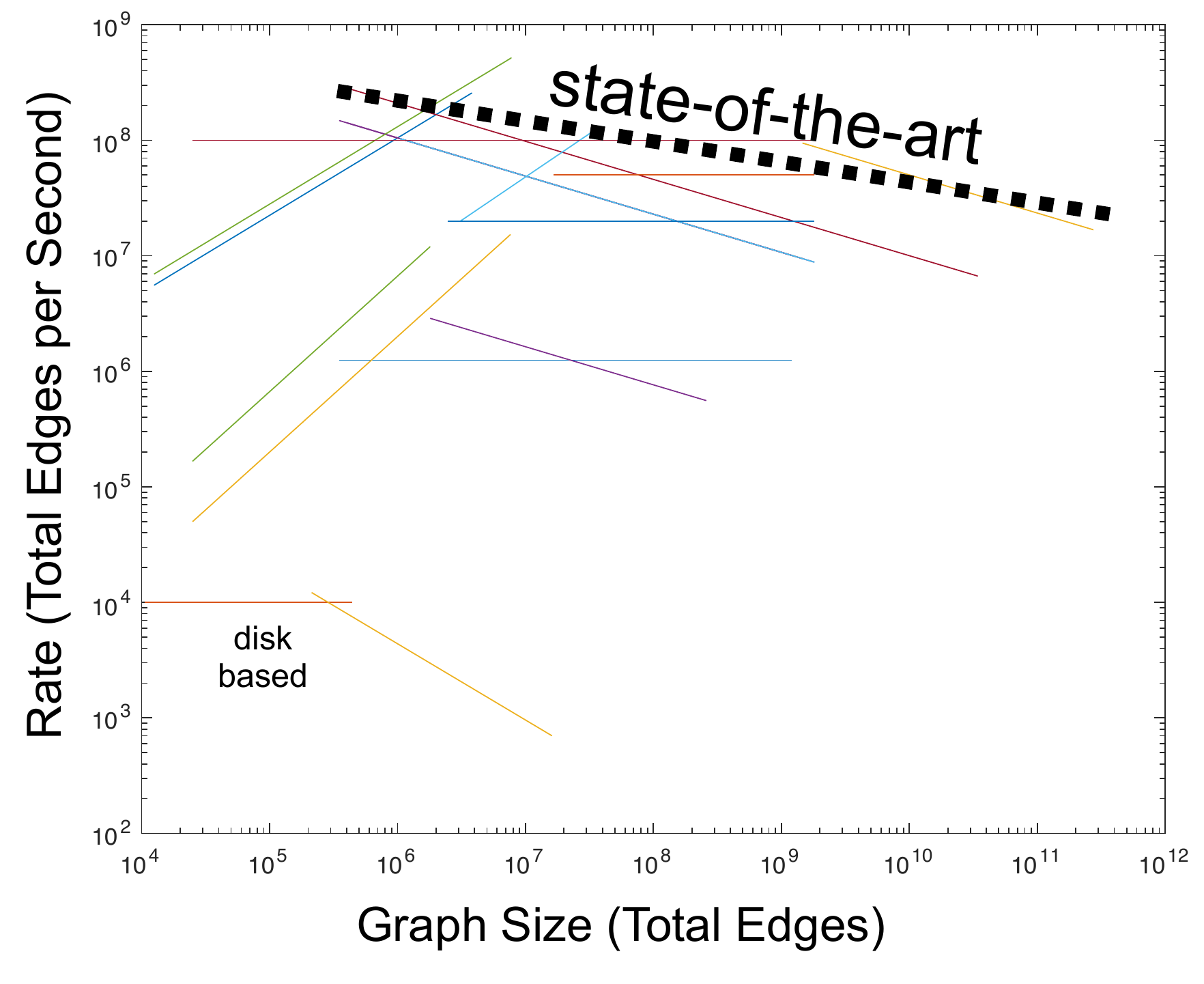}
\caption{Model fits of triangle execution rate vs. number edges for selected Graph Challenge 2017 triangle counting submissions.  State-of-the-art is denoted by the black dashed line. The one disk-based submissions shows the importance of large, high-speed memories for this application.}
\label{fig:RatePerformance}
\end{figure}

  The model fits in Figures~\ref{fig:Champions}, \ref{fig:Finalists}, and \ref{fig:HonorableMentions} can be put in a more understandable form by transforming them to the normalized equation
$$
   T_{\rm tri} = (N_e/N_1)^\beta
$$
where $N_1 = \alpha^{-1/\beta}$ is the number edges that take 1 second to process.   The normalized parameters $N_1$ and $\beta$, along with the largest values of $N_e$, are shown in  Table~\ref{table:NormalizeModel} for  each submission.  Submissions with larger $N_e$, larger $N_1$, and smaller $\beta$ perform best.  Combined, these suggest a current state of the art performance model of
$$
   T_{\rm tri} = (N_e/10^8)^{4/3}
$$
and
$$
   {\rm Rate} = \frac{10^8}{(N_e/10^8)^{1/3}}
$$
The current state-of-the-art can also be seen by plotting all the model fits $T_{\rm tri}$ together (see Figures~\ref{fig:TimePerformance} and \ref{fig:RatePerformance}).  Given the enormous diversity in processors, algorithms, and software, this relatively consistent picture of the state-of-the-art suggests that the current limitations are set by common elements across these benchmarks, such as memory bandwidth.

\begin{table}
\caption{{\rm Triangle counting time model fit coefficients for $T_{\rm tri} = (N_e/N_1)^\beta$ for large values of  $N_e$.}}
\centering
\begin{tabular}{lllcc}
\hline
Ref & Submission & max $N_e$ & $N_1$ & $\beta$ \\
\hline
\cite{bisson2017static}          & Bisson-Nvidia-2017       & $1.5\times10^9$ & $3\times10^7$ & $4/3$ \\
\cite{pearce2017triangle}        & Pearce-LLNL-2017         & $2.7\times10^{11}$ & $2\times10^8$ & $4/3$ \\
\cite{voegele2017parallel}       & Voegele-UTAustin-2017    & $1.8\times10^9$ & $3\times10^7$ & $4/3$ \\
\cite{wolf2017fast}              & Wolf-Sandia-2017         & $1.8\times10^9$ & $3\times10^7$ & $4/3$ \\
\cite{hu2017trix}                & Hu-GWU-2017              & $3.4\times10^{10}$ & $5\times10^7$ & $4/3$ \\
\cite{smith2017truss}            & Smith-UMN-2017           & $1.2\times10^9$ & $1\times10^6$ & $1$ \\
\cite{tom2017exploring}          & Tom-UMN-2017             & $1.8\times10^9$ & $5\times10^7$ & $1$ \\
\cite{date2017collaborative}     & Date-UIUC-2017           & $2.6\times10^8$ & $3\times10^6$ & $4/3$ \\
\cite{hutchison2017distributed}  & Hutchison-UWash-2017     & $1.6\times10^7$ & $3\times10^4$ & $5/3$ \\
\cite{low2017first}              & Low-CMU-2017             & $1.8\times10^9$ & $1\times10^8$ & $1$ \\
\cite{mowlaei2017triangle}       & Mowlaei-UPitt-2017       & $1.8\times10^9$ & $5\times10^7$ & $1$ \\
\hline
\end{tabular}
\label{table:NormalizeModel}
\end{table}

\section{Conclusion}

The rapid increase in the use of graphs and has inspired  new ways to measure and compare
the attributes of graph analytic systems. The MIT/Amazon/IEEE Graph Challenge was created to stimulate research  in graph analysis software, hardware, algorithms, and systems.   The GraphChallenge.org website makes available to the world many pre-processed graph data sets, graph generators, graph algorithms, prototype serial implementations in a several languages, and defined metrics for assessing performance.  In 2017, Graph Challenge received 22 submissions by 111 co-authors from 36 organizations.  The submissions covered new developments in hardware, software, algorithms, systems, and visualization.  The comparable measurements provided by these submissions are an invaluable resource for determining the state-of-the-art in graph processing.  Triangle counting was the most popular submission and produced over 350 distinct measurements of triangle execution time on a wide variety of graphs.  Analysis of each of these submissions shows that execution time is a strong function of the number of edges in the graph, $N_e$, and is typically proportional to $N_e^{4/3}$ for large values of $N_e$.  Combining the model fits of each of the submissions presents a picture of the current state-of-the-art of graph analysis, which is typically $10^8$ edges processed per second for graphs with $10^8$ edges. These results are $30$ times faster than serial implementations commonly used by many graph analysts and reinforces the importance of making these performance benefits available to the broader community.

Graph Challenge provides a clear picture of current graph analysis systems and underscores the need for new innovations to achieve high performance on very large graphs.  In the future, Graph Challenge will examine additional data sets, algorithms, and implementations.  One area of interest is sparse neural networks. Neural networks are becoming widely used in many disciplines.  In some contexts, sparse neural networks are limited by their size.  One approach to alleviating these size limitations is to make the representation of a neural network sparse, which may require new systems to effectively perform neural network computations.


\section*{Acknowledgments}
%
%

The authors wish to acknowledge the following individuals for their contributions and support: Alan Edelman, Charles Leiserson, Steve Pritchard, Michael Wright, Bob Bond, Dave Martinez, Sterling Foster, Paul Burkhardt, and Victor Roytburd.



\bibliographystyle{IEEEtran}
\bibliography{aarabib}
%

\end{document}